# DISCUSSION OF: STATISTICAL ANALYSIS OF AN ARCHEOLOGICAL FIND

By Holger Höfling[1] and Larry Wasserman[2]

*Stanford University and Carnegie Mellon University*

*There are no small coincidences and big coincidences! There are only coincidences!*

From "The Statue" episode of *Seinfeld*.

**1. Introduction.** Andrey Feuerverger has undertaken a serious challenge. The subject matter is controversial and finding a sensible way to formulate the problem in a rigorous statistical manner is difficult.

The paper is notable for its thoroughness. We have rarely seen a paper on an applied problem that provides so much background material. Most importantly, the author is very careful to document all his assumptions and to remind the reader that the conclusion is sensitive to these assumptions. He resists the temptation to present his results in a sensationalistic way. Rather, he conveys his analysis in a dispassionate, understated tone. Nonetheless, he could still end up on *Oprah*.

We are trying to assess the probability of a hypothesis when the hypothesis is formed after seeing the data. This is a notoriously difficult problem. As Feuerverger notes, coincidences are common. But just how common?

One response—the nihilistic approach—is to say that it is impossible and stop there. We have much sympathy with the nihilists in a problem like this. Perhaps the scientifically honorable path is to say that any answer is misleading so it is better to provide no answer. But ultimately this is unsatisfying and we accept the author's approach to provide an analysis with many caveats.

The question may be framed formally as follows. We observe an outcome $x$—a tomb with interesting names—and we want to know: is this outcome

Received September 2007.

[1]Supported by an Albion Walter Hewlett Stanford Graduate Fellowship.

[2]Supported by NSF Grant CCF-06-25879. Thanks to Rob Tibshirani and Isa Verdinelli for helpful comments.

surprising? One way to quantify surprisingness is to perform the following steps:

1. Construct a sample space $\mathcal{X}$ that contains $x$.
2. Identify all the outcomes $A$ that would have been considered surprising if they had been observed.
3. Construct an appropriate null distribution $P_0$.
4. Compute the $p$-value $p = P_0(A)$.

The most difficult step is identifying the set $A$ of interesting outcomes. It is explicitly counterfactual to ask if an outcome would have been surprising if it had occurred, knowing that it did not occur.

**2. Feuerverger's approach.** What the author has proposed is both interesting and reasonable. Numerous judgement calls have to be made but they have been carefully documented. Our summary of Feuerverger's method is this: The sample space is chosen to be sets of names on ossuaries, subject to some restrictions. The null measure is essentially random sampling from an onomasticon. The author defines a statistic (RR) that maps sets of names into products of numbers. These numbers are essentially sample proportions, modified to take into account various nuances such as surprisingness of versions of names. The result is a very small $p$-value suggesting that the find is indeed surprising.

The 'Mariamenou $\eta$ Mara' inscription has a very big effect on Feuerverger's RR statistics. An explanation for this is that the RR statistic becomes more significant if broad name categories are being subdivided into special name renditions, even if the particular name renditions are not relevant. The following example illustrates this point:

> A population has three names $A$, $B$ and $C$ each with frequency 1/3. $A$ has 2 name renditions $A_1$ (1/3 of $A$) and $A_2$ (2/3 of $A$). Our family has two members named $A$ and $B$, and $A_1$ and $A_2$ are both relevant. The uncovered tomb has one inscription $A_1$. When only considering broad name categories, we have $RR(A) = 1/3$, $RR(B) = 1/3$ and $RR(C) = 0$. When the null is random drawing from the population, the $p$-value is then 2/3.
>
> When taking name renditions into account, $RR(A_1) = 1/9$, $RR(A_2) = 2/9$, $RR(B) = 1/3$ and $RR(C) = 0$ giving $p$-value of 1/9. The $p$-value decreased although both name renditions were considered relevant. The change in $p$-value can be even more substantial in more complicated cases.

In this comment, we present a Frequentist and a Bayesian approach that do not have this problem and yield quite different results.

**3. A different approach.** We would like to consider a different way of defining the basic event $A$. Our approach is more expansive and, as a result, more conservative. Instead of asking "What is the probability of getting this set of names?" we ask "What is the probability of getting some interesting set of names if one looks at several tombs?"

Let $\mathcal{X}$ be all name sets. Examples of sample points in $\mathcal{X}$ are

$$x = \{\text{Salome}\},$$

$$x = \{\text{Levi, Hanan, Simon, Mariam}\},$$

$$x = \{\text{Joseph, Jesus, Sarah}\},$$

and so on. Define a list of target names $S$. The list should include all names that will spark interest. We take this to be either the big set

$$S = \{\text{Mariam, Mary, Salome, James, Joseph, Joanna, Martha}\}$$

or the small set

$$S = \{\text{Mariam, Mary, Salome, James, Joseph}\}.$$

The name "Jesus" is not included because we will treat it separately. We assume that a tomb would have triggered interest if its name set $B$ has sufficient overlap with $S$. We lump together different version of names since interested observers would surely argue that a tomb is interesting if there is any way at all of matching the found names to potentially interesting names. Denote the name sets in the tombs by $B_1, \ldots, B_N$. Say that $B_i$ is interesting if

$$|B_i \cap S| \geq 3 \quad \text{and} \quad \text{"Jesus"} \in B_i.$$

We denote the probability of this event by $\pi_i$. Assuming independence of name assignments in and across tombs, the $p$-value is

$$p = 1 - \prod_{i=1}^{N}(1 - q(n_i, \pi_i))$$

where $n_i$ is the number of ossuaries in tomb $B_i$,

$$q(n_i, \pi) = p_J \mathbb{P}(Y_i \geq 3), \qquad Y_i \sim \text{Binomial}(n_i - 1, \nu),$$

$\nu$ is the probability that a single name drawn at random is in $S$ and $\pi_J$ is the probability of drawing the name "Jesus." We do not take $\pi_J$ to be the probability of drawing "Jesus son of Joseph" because the tomb could have been considered interesting if it had only said "Jesus."

For our calculations we take $N = 100$, $n_i = 6$. The number 100 comes from the fact that there are about 1000 tombs but only 10 percent have been excavated. Hence $\pi_i = \pi$ does not vary with $i$. We consider two possibilities for the male-femail ratio: (i) equal or (ii) unequal as represented by the onomasticon. For example, in case $S$ is equal to the first (big) choice, the male/female ratio is equal we get

$$\nu = \frac{1}{2}\left(\frac{231 + 103 + 45}{2509}\right) + \frac{1}{2}\left(\frac{81 + 63 + 21 + 12}{317}\right) = 0.3547.$$

The value of $\pi$ and $p$ for the different combinations of assumptions is as follows:

| $S$ | m/f ratio | $\pi$ | $p$-value |
|-----|-----------|-------|-----------|
| big | equal | 0.005 | 0.393 |
| big | not equal | 0.002 | 0.183 |
| small | equal | 0.003 | 0.290 |
| small | not equal | 0.002 | 0.158 |

We reiterate that we have not treated name variations as special. But the calculation is invariant under splitting names into subcategories since we are finding the probability of a set of interesting names, not a particular name. We also ignored family structure. We now consider two variations. We consider replacing "Jesus" with "Jesus son of Joseph" by multiplying these two probabilities. We also consider taking $N = 1000$ to reflect the unobserved tombs. The results are:

|                    | $N = 100$ | $N = 1000$ |
|--------------------|-----------|------------|
| Jesus              | 0.16      | 0.82       |
| Jesus son of Joseph | 0.01     | 0.13       |

There is one case where the $p$-value is small. But the lack of robustness of this result does not make us confident in reporting a small $p$-value.

We conclude that the observed event is not rare at all. The chance that an observer would find a tomb that could be said to contain interesting target names is large. This is due to the fact that the interesting names are common and that the many tombs provide many opportunities for apparent surprises.

**4. Bayesian analysis.** Now we consider a Bayesian analysis of the problem. We need to compute

$$P(\theta = 1|x) = \frac{P(x|\theta = 1)P(\theta = 1)}{P(x|\theta = 1)P(\theta = 1) + P(x|\theta = 0)P(\theta = 0)},$$

where $x$ denotes the data, $\theta = 1$ that the tomb is from the NT family and $\theta = 0$ that the tomb is from the normal population.

In the frequentist approach, a partial ordering has to be defined on the space of all outcomes. Feuerverger does this using the RR statistic and the approach described above uses intersection of name sets. However, discerning the exact ordering on the space of outcomes may be hard or people might not agree with it. The advantage of the Bayesian approach is that the alternative distribution only has to be defined at the point $x$ and no ordering on the space of possible outcomes is needed.

4.1. *Posterior probability.* Let us introduce a little more notation at this point. Let $c$ be the configuration of a tomb, $g$ be its genealogy, $n = (n_1, \ldots, n_K)$ the broad name categories and $r = (r_1, \ldots, r_K)$ the particular name renditions. Assuming that every name rendition only depends on $\theta$ and its broad name category, we can write

$$P(x|\theta) = P(c, g|\theta) P(n|g, c, \theta) \prod_{i=1}^{K} P(r_i|n_i, \theta).$$

**Simplifying assumptions:** To make the computations easier, we make two more assumptions:

1. The configuration and genealogy we expect to see in the NT family tomb is not different from the rest of the population, that is, $P(c, g|\theta = NT) = P(c, g|\theta = P)$.
2. The particular name renditions we expect to see in the NT family tomb are no different than what we expect to see in the rest of the population, that is, $P(r_i|n_i, \theta = NT) = P(r_i|n_i, \theta = P)$. This assumption will be relaxed later.

Then the posterior odds are

$$\frac{P(\theta = 1|x)}{P(\theta = 0|x)} = \frac{P(\theta = 1)}{P(\theta = 0)} \cdot \frac{P(n|c, g, \theta = 1)}{P(n|c, g, \theta = 0)}.$$

4.2. *Distributions.* First, we define the prior distribution. Feuerverger estimates the number of tombs in the area to be about $N = 1100$. Also, let the prior probability of the NT family having a tomb at all be $t$. Then

$$P(\theta = 1) = t\frac{1}{N}, \qquad P(\theta = 0) = 1 - P(\theta = NT).$$

In order to be optimistic, we take $t = 1$ and get prior odds of

$$\frac{P(\theta = 1)}{P(\theta = 0)} = \frac{1}{1099}.$$

This prior can be thought of as a Bayesian approach to account for data snooping, that is, the potential to searching through many tombs.

For the null distribution, names are drawn randomly using the name frequencies in Ilan. Men and women are being treated separately and the list of names $n$ is treated as unordered.

When specifying the probability distribution under the alternative, it is necessary to weigh flexibility against complexity. Here we want to take the

TABLE 1
*Weights for each of the persons listed*

| Scenario | Jesus son of Joseph | James | Joses | Matthew | Judas | Others |
|---|---|---|---|---|---|---|
| Neutral | 20 | 3 | 3 | 62/2509 | 171/2509 | 3 |
| Optimistic | ∞ | 1 | 1 | 62/2509 | 171/2509 | 0 |
| Scenario | Marya (mother) | Mariam (sister) | Salome (sister) | Mary Magdalene | | Others |
| Neutral | 10 | 3 | 3 | 3 | | 3 |
| Optimistic | ∞ | 1 | 1 | 0 | | 0 |

Drawing from the set is being done with probabilities proportional to the weights without replacement. The weight in the "others" category is the weight for all not listed persons.

following approach: Specify a set of names from the NT family (separately for men and women) and assign each name a weight as to how likely it is to find this person in the NT family tomb. Then, the probability of a specific tomb is calculated by drawing from the nameset without replacement according to the weights. The weights can be determined in an optimistic or more conservative fashion (see Table 1).

For simplicity, the probability of being in the generational ossuary is taken to be the same for everyone, under the null as well as the alternative.[1]

**Neutral scenario:** In this case, we chose the weights in a fashion that seemed reasonable to us when we do not consider the information gathered from the tomb. Also, each name in the tomb is taken as its broad name category and it is assumed that no additional information for special name renditions is available for the NT family.

**Neutral with special renditions:** Here, we use the same weights as in the neutral scenario, however account for the special "Mariamenou $\eta$ Mara" rendition. Each of the other inscriptions on the ossuaries is not special, so we do not make any adjustments for those. A priori, we could not have known the inscription "Mariamenou $\eta$ Mara," so how do we account for it? Under $\theta = P$, we assume that for the Marya name category, the probability of seeing a new previously unseen name is 1/80. For $\theta = NT$, we assume that special name renditions are more likely, say 1/10. Assuming that 'Mariamenou $\eta$ Mara' could in some way be interpreted for Maria (mother), Mariam (sister) and Maria Magdalena, this raises the odds by a factor 8 over the neutral scenario.

**Optimistic scenario:** We also wanted to explore the effect of having very optimistic assumptions which are to a large degree influenced by what has been observed in the tomb. Jesus and his mother are taken to be in the tomb for sure. For the rest of the men, the weights are equal for both brothers and set to the normal name frequency in the population for Matthew and Judas. The overall effect of this choice of weights is to effectively ignore the Matthew and Judas ossuaries, assume that one of the ossuaries is from a brother and one from a sister of Jesus and assign all eligible brothers and sisters the same weight.

---

[1] This may be viewed as an oversimplification, however as the weights provide ample opportunity to fine tune prior beliefs, we do not see this as practically important.

TABLE 2
*Posterior probability that the Talpiyot tomb
belongs to the NT family under various scenarios*

| Scenario | Probability |
|---|---|
| Neutral | 3.4% |
| Neutral—special renditions | 21.8% |
| Very optimistic | 64.1% |

4.3. *Results.* Even in the optimistic scenario, there is only about a 60% chance of the tomb belonging to the NT family. In the other two, more realistic schemes, the probability is only 22% and 3% (see Table 2). Just as Feuerverger, we also did not consider the generational part of the "Judas, son of Jesus" ossuary. Including it in the analysis would be possible; however, as prior beliefs about a possible son of Jesus are very strong, this may have overwhelmed the rest of the analysis and therefore we decided to exclude it.

**5. Conclusion.** When asked to analyze these data, we suspect that many statisticians would have said that the problem is too vague and would have stopped there. We commend Andrey Feurverger for plunging in and doing a serious analysis. Our analysis suggests that the finding does not lend support to the hypothesis that the find is indeed the tomb of the NT family. Ultimately, scholars of history and archeology will judge the validity of the claims about this find.

DEPARTMENT OF STATISTICS
STANFORD UNIVERSITY
STANFORD, CALIFORNIA 94305
USA
E-MAIL: hhoeflin@stanford.edu

DEPARTMENT OF STATISTICS
CARNEGIE MELLON UNIVERSITY
PITTSBURGH, PENNSYLVANIA 15213
USA
E-MAIL: larry@stat.cmu.edu